\documentclass[manuscript]{aastex}

\slugcomment{Not to appear in Nonlearned J., 45.}

\shortauthors{White et al.}

\usepackage[margin=1in]{geometry} 
\usepackage{amsmath,amsthm,amssymb}
\usepackage{booktabs}
\usepackage{subfigure}
\usepackage{footnote}
\usepackage{ulem}
\usepackage{color}
\usepackage{graphicx}
\usepackage{times} 
\usepackage{amssymb}
\usepackage{amsmath}
\usepackage{lscape}
\usepackage{url}
\usepackage{multirow}
\usepackage{longtable}
\usepackage[english]{babel}
\usepackage{float}

 \graphicspath{{./../files/}}

\begin{document}


\title{ALMA Observations of HD141569's Circumstellar Disk}

\author{J.A. White\altaffilmark{1}}

\email{jawhite@astro.ubc.ca}

\author{A.C. Boley\altaffilmark{1}}

\author{A.M. Hughes\altaffilmark{2}}
\author{K.M. Flaherty\altaffilmark{2}}
\author{E. Ford\altaffilmark{3,4,5}}
\author{D. Wilner\altaffilmark{6}}
\author{S. Corder\altaffilmark{7}}
\author{M. Payne\altaffilmark{6}}

\altaffiltext{1}{Department of Physics and Astronomy, 6224 Agricultural Road, Vancouver, BC V6T 1Z1}
\altaffiltext{2}{Department of Astronomy, Van Vleck Observatory, Wesleyan University, 96 Foss Hill Drive, Middletown, CT 06459, USA}
\altaffiltext{3}{Center for Exoplanets and Habitable Worlds, 525 Davey Laboratory, The Pennsylvania State University, University Park, PA, 16802-2111, USA}
\altaffiltext{4}{Department of Astronomy and Astrophysics, The Pennsylvania State University, 525 Davey Laboratory, University Park, PA 16802-2111, USA}
\altaffiltext{5}{Center for Astrostatistics, The Pennsylvania State University, 417C Thomas Building, University Park, PA 16802-2111, USA}
\altaffiltext{6}{Harvard Smithsonian Center for Astrophysics, 60 Garden St, Cambridge, MA 02138, USA}
\altaffiltext{7}{North American ALMA Science Center, National Radio Astronomy Observatory, 520 Edgemont Road, Charlottesville, VA, 22903, USA}


\begin{abstract}
We present ALMA band 7 (345 GHz) continuum and $^{12}$CO(J = 3-2) observations of the circumstellar disk surrounding HD141569. At an age of about 5 Myr, the disk has a complex morphology that may be best interpreted as a nascent debris system with gas. Our  $870\rm~\mu m$ ALMA continuum observations resolve a dust disk out to approximately $ 56 ~\rm au$ from the star (assuming a distance of 116 pc) with $0."38$ resolution and $0.07 ~ \rm mJy~beam^{-1}$ sensitivity. We measure a continuum flux density for this inner material of $3.8 \pm 0.4 ~ \rm mJy$ (including calibration uncertainties). The $^{12}$CO(3-2) gas is resolved kinematically and spatially from about 30 to 210 au. The integrated $^{12}$CO(3-2) line flux density is $15.7 \pm 1.6~\rm Jy~km~s^{-1}$. We estimate the mass of the millimeter debris and $^{12}$CO(3-2) gas to be $\gtrsim0.04~\rm M_{\oplus}$ and $\sim2\times 10^{-3}~\rm M_{\oplus}$, respectively. If the millimeter grains are part of a collisional cascade, then we infer that the inner disk ($<50$ au) has $\sim 160~\rm M_{\oplus}$ contained within objects less than 50 km in radius, depending on the planetesimal size distribution and density assumptions. MCMC modeling of the system reveals a disk morphology with an inclination of $53.4^{\circ}$ centered around a $\rm M=2.39~ M_{\odot}$ host star ($\rm Msin(i)=1.92~ M_{\odot}$). We discuss whether the gas in HD141569's disk may be second generation.  If it is, the system can be used to study the clearing stages of planet formation.

\end{abstract}


\keywords{circumstellar matter – stars: individual (HD 141569)}

\section{Introduction}

While many details of planet formation are not fully understood \citep{johansen, raymond14, chabrier, helled}, significant debris is expected to be produced by the planet-building process. These leftovers, such as asteroids and comets, dynamically and collisionally evolve over a planetary system's lifetime, creating a steady source of dust and small grains, which would otherwise be depleted on short timescales \citep{matthews}. Thus, the presence of circumstellar debris around a star is taken as evidence that planet building was at least partially successful in that system. When debris structures are resolved, the morphologies can be used to place constraints on the architecture of putative planets \citep{kuchner,quillen, moro, stark} and to potentially understand the dynamical history of a system \citep{raymond}. Multi-frequency observations can further be used to constrain dust properties \citep{wyatt}, giving a way to explore the debris itself.

Among known debris disks, a limited number contain gas, as detected in radio molecular line emission. This includes $\beta$ Pic \citep{zuckerman95, dent}, HD 131835 \citep{moor15}, HD21997 \citep{moor11, moor13}, and 49 Cet \citep{hughesa} with estimated ages of 12 Myr, 16 Myr, 30 Myr, and 40 Myr, respectively. These systems are older than the typical lifetimes of gaseous disks, as inferred from IR excess and accretion \citep[e.g.,][]{mamajek}. Furthermore,  if the gas has a primordial origin (i.e., from the formation of disk itself), the gas abundances need to be reconciled with photoevaporation rates \citep{alexander}  and CO photodissociation timescales \citep{vandishoeck_black_1988, visser}. Photoevaporation rates may not be constant throughout the lifetimes of the disk, and the radial distribution of gas is influenced by both UV and X-ray sources \citep[e.g.,][]{gorti}.

Instead of primordial, the gas could be second-generation, produced by the early evolution of a comet reservoir \citep{dent} through impact vaporization or sublimation of impact-generated particulates. It nonetheless remains unclear whether there is sufficient mass in comets to explain the amount of gas detected in these systems \citep{matthews, moor13}. Regardless of the reason, the existence of this gas has implications for planet building. For example, while the measured gas masses are too small to contribute significantly to gas giant planet formation, the gas could still contribute to planetary atmospheres and potentially, for high enough gas masses, continue to affect small-grain dust.

If the gas does have a debris origin, then the relative debris and gas morphologies, along with dynamical models of the system, can be used to probe the clearing stages of planet formation and serve as a probe of disk mass during that evolutionary stage. As such, debris+gas systems can potentially offer significant constraints on planet formation theory \citep{kospal, wyatt_stages}. To this end, HD141569 is of particular interest.

HD141569 is a B9.5 Ve star at a distance\footnote{\citet{hipparcos} find a distance of $99\pm8$ pc, whereas the re-analysis of the {\it Hipparcos} data yields a distance of $116\pm8$ pc \citep{van}. Throughout the literature, both distances are used for HD141569.   In this manuscript, when reporting linear sizes from other work, we simply use their reported values.  For the stellar, dust, and gas masses that we derive here, we will discuss how the results are expected to scale with distance.} of $116\pm8$ pc \citep{van}.  At an age of about 5 Myr, it is surrounded by a complex dust and gas disk \citep{vandenancker, weinberger, fisher}. At distances $>100$ au from the star, large-scale spiral structure has been detected in optical scattered light, revealing at least two well-defined ring/spiral-like structures \citep{weinberger, clampin}. One spiral is between $\sim 175$ and 210 au, and the other between $\sim 300$ and 400 au. The rings/spirals are bright, with an optical depth $\sim 0.01$ in the outer arm \citep{clampin} and a scattered light flux density of $4.5\pm0.5 ~\rm mJy$ at $1.6 ~\mu$m  \citep{mouillet, augereau}.

In addition to having a large extended disk, HD141569 also hosts an inner dust system. This disk was first detected by excess emission in the mid-infrared using IRAS \citep{walker, andrillat}. Observations at 12, 25, 60, and 100 $\mu$m wavelengths \citep{walker} led to a calculated disk radius of 47-63 au, based on modeling \citep[see also]{fisher, marsh}. \citet{thi} used archival VLT data at 8.6 $\mu$m to  resolve the inner system out to $\sim50$ au. SED modeling suggests that the inner edge of small grains must be at about 10 au with a likely peak at 15 au \citep{malfait, maaskant}. \textbf{Select} previous continuum observations are summarized in Table 1.

If the dust's origin is debris, HD141569 may be viewed as the youngest of the gas-rich debris systems.  By ``debris'', we mean that the majority of the (sub)millimeter emission from solids is associated with grains that have already been incorporated into a parent body and re-released into the nebula.  If the solids have not already been processed into parent bodies, then they reflect the initial growth stages of grains in planet-forming disks.

\begin{table}[H]
\begin{center}
\caption{Summary of select previous HD 141569 debris disk observations. Uncertainties provided when available. References listed are: (1) \citet{fisher}; (2) \citet{walker}; (3) \citet{marsh}, (4) \citet{mouillet}; (5) \citet{augereau}; (6) \citet{nilsson}; (7); \citet{sylvester01}}
\label{my-label}
\begin{tabular}{ l|l|l|l|l }

\hline\hline
\textbf{Features} & \textbf{Wavelength [$\mu$m]} & \textbf{Flux Density [Jy]} & \textbf{Instrument} & \textbf{Ref.} \\
\hline
     Continuum & $10.8$  & $0.318 \pm 0.016$ & Keck OSCIR & (1) \\
     Continuum & $18.2$  & $0.646 \pm 0.035$ & Keck OSCIR & (1) \\
     Continuum & $12, 25, 60, 100$  & $0.66, 1.99, 5.37, 3.34$ & IRAS & (2) \\
     Continuum & $12.5, 17.9, 20.8$  & $0.333, 0.936, 1.19$  & KECK MIRLIN & (3)\\
     & & $\pm .022, \pm, .094, \pm 0.16$ & \\
     Spiral Structure   & $1.6$  &  $0.0045 \pm 0.0005$   &         HST      & (4,5) \\
     Total System  & $870$ & $0.0126 \pm 0.0046$  & APEX & (6) \\
     Total System & $1350$ & $0.0054 \pm 0.001$ & JCMT SCUBA & (7) 

\label{past}
\end{tabular}
\end{center}
\end{table}

The total gas mass has been constrained to be roughly between 13 and 200 M$_{\oplus}$ \citep{zuckerman95, thi, flaherty}, depending on assumed abundance ratios and model fitting. Most of this mass is likely located in the outer system, where CO kinematics suggest that the gas is non-uniformly distributed in radius. Tracers of hot gas such as ro-vibrational CO lines in the near-infrared \citep{brittain02, goto} show that there is a region of tenuous CO gas distributed between 10 and at least 50 au, seemingly commensurate with the inner dust system.

HD141569 may be in a stage where the outer gas regions have, at least in part, a primordial component, but the inner region associated with millimeter grains may arise from the collisional evolution of parent bodies. We must also entertain whether the outer gaseous disk is dominated by second-generation gas, making the entire system an early-stage debris disk. 

In this paper we present ALMA band 7 observations of the inner dust and outer gas systems. Section 2 is an overview of the observations and data reduction.  The $870 \mu m$ continuum and $^{12}$CO(J = 3-2) (hereafter CO(3-2)) spectral imaging and analysis of the gas disk are given in section 3. We describe mass calculations and discuss interpretations in Section 4.  Section 5 summarizes the results.

\section{Observations}

The data were acquired on 21 May 2014 as part of the ALMA cycle 1 campaign (project ID 2012.1.00698.S). Observations were made in two execution blocks (EBs), but one EB could not be calibrated due to phase amplitude and water vapor radiometer (WVR) problems. The total integration time for the successful EB was 1.43 hr (0.79 hr on target). A compact configuration was used with 32 antennas; the longest baseline was 650.3 m.  Observations were centered on HD141569  using J2000 coordinates RA = 15 hr 49 min 57.73 sec  and $\delta = -3^{\circ} 55' 16.62''$. 

To acquire high S/N data in both continuum and CO(3-2) efficiently, observations were taken in band 7 (at $\sim345$ GHz) with the correlator setup using the Frequency Division Mode (FDM) and dual polarization. Four different spectral windows were used with 1875 MHz bandpasses at rest frequency centers of 335, 337, 345, and 347 GHz. These locations were chosen to maximize continuum sensitivity while also overlapping the CO(3-2) transition. The correlator in FDM gives 3840 channels of width 488 kHz, which corresponds to a velocity resolution of $0.85$ km s$^{-1}$.

Titan and quasar J1550+0527 were used for absolute flux and bandpass calibration, respectively.   Atmospheric variations at each antenna were monitored continuously using the WVRs. The estimated WVR thermal contribution to path fluctuations is $5.8$ $\mu$m per antenna.

Data were reduced using the Common Astronomy Software Applications (CASA) package \citep{casa_reference}. Antenna 14 was flagged during quality assurance (QA), leaving 31 antennas for the final data product. In addition, spectral windows 1 and 3 each exhibited 120 bad channels (1/32 of the bandwidth), which were also flagged. Antenna 14 and the flagged channels were removed from the data prior to reduction and subsequent analyses using the task \textit{split}. The data reduction in CASA included WVR calibration; system temperature corrections; and bandpass, flux, and phase calibrations with Titan and quasar J1550+0527.

\section{Results}

Table \ref{data} summarizes observed system properties for both the dust and gas. The continuum flux density is determined by fitting a disk model to visibilities (see Sec.\,3.1), while the gas flux density is taken from integrating within the $3\sigma$ contours of the zeroth moment maps (see Sec.\,3.2).

The peak intensity and angular size are taken from the CLEANed images, assuming a distance of $116~\rm pc$ for linear scales. The uncertainties for the flux densities and the line fluxes include the $\sigma_{\rm{RMS}}$ of the observations and an absolute flux calibration uncertainty of $\sim10\%$ added in quadrature.  The uncertainties in the intensities only include the  $\sigma_{\rm{RMS}}$.

\begin{table}[H]
\caption{Summary of observed values and  for both gas and dust. The flux densities are determined by fitting the visibilities by a disk model (see Sec.\,5.1). The peak intensity and angular size are derived from the CLEANed images. Linear sizes assume a distance of $116$ pc and are measured across the semimajor axis of the Continuum and gas.  The uncertainties for the flux densities and the line fluxes include the $\sigma_{\rm{RMS}}$ of the observations and an absolute flux calibration uncertainty of $\sim10\%$ added in quadrature.  The uncertainties in the intensities only include the $\sigma_{\rm{RMS}}$.}

\centering 
\begin{tabular}{c | c | c} 
\hline\hline 

    Parameter & Continuum [Debris] & Gas [CO 3-2] \\
	\hline
    Flux Density & $3.8 \pm 0.4 $ mJy             & $15.7 \pm 1.6 $ Jy km s$^{-1}$ \\
    Peak Intensity    & $1.74 \pm 0.24 $ mJy beam$^{-1}$ & $0.90 \pm 0.16 $ Jy beam$^{-1}$  \\
    Angular Radius & $0."49$ ($\sim 56$ au)         & $1.8"$ ($\sim 210$ au)  \\    
    $\sigma_{\rm{RMS}}$ & $0.070$ mJy beam$^{-1}$   & $0.028$ Jy beam$^{-1}$   \\
     Synthesized Beam Area & $0.163$ arcsec$^{2}$   & $0.121$ arcsec$^{2}$   \\    
     Beam major axis FWHM & $0."42$ & $0."34$ \\
     Beam major axis FWHM & $0."34$ & $0."31$ \\
     Beam Position Angle (PA) & $-61.1^{\circ}$& $-77.1^{\circ}$

\label{data}
\end{tabular}
\end{table}
\subsection{Continuum}

\begin{figure}[H]
\centering
\includegraphics[width=\textwidth]{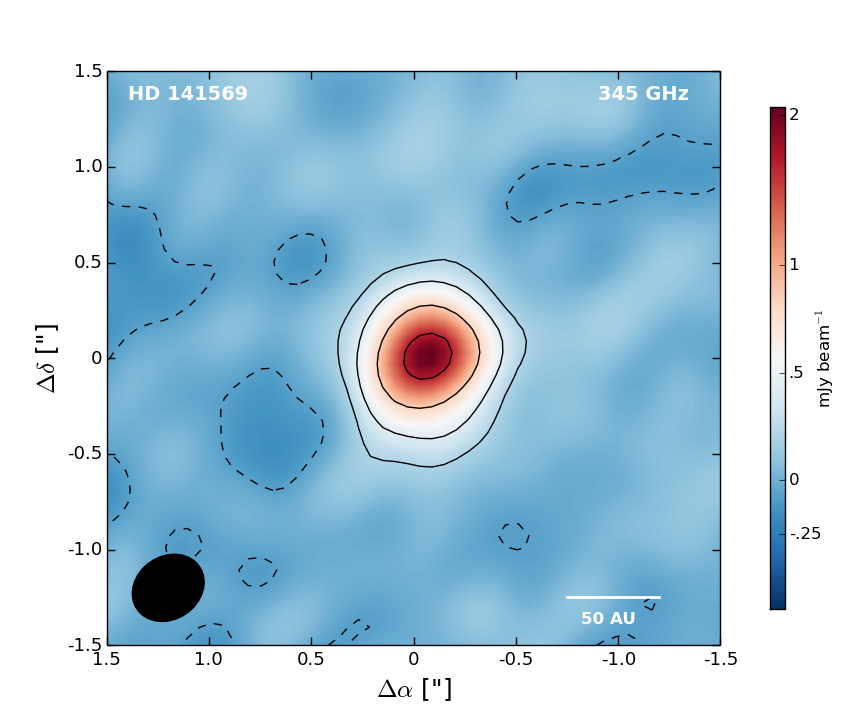}
\caption{CLEANed $870~ \mu m$ continuum image of HD 141569. The contours represent $3, 6, 12 $ and $ 21 \times \sigma_{RMS}$ noise ($\sigma_{RMS} = 0.070~\rm mJy~beam^{-1}$). The dashed contour represents $-1\sigma$. The solid ellipse in the bottom left represents the beam size. A $50$ au scale (assuming a system distance of $116$ pc) is given in the bottom right. The peak intensity is $1.74 \pm 0.24 $ mJy beam$^{-1}$. Coordinates are given as offset from the phase center. North is up and East is to the left.
\label{clean}}
\end{figure}

The dust emission is clearly resolved by the ALMA beam. The continuum (with the CO channels removed) is deconvolved and imaged using CASA's CLEAN algorithm. The average wavelength across the frequency range is $870 \mu m$. A threshold of $\frac{1}{2} \times \sigma_{RMS}$ and a natural weighting are used to produce the final cleaned product in Fig. \ref{clean} (with contours corresponding to $3, 6, 12 $ and $ 21 \times \sigma_{RMS}$). The inner disk around HD 141569 is imaged out to 56 au (assuming a distance of 116 pc). The longest baseline is unable to resolve a central clearing of $< 15\rm~au$, leading to a central peak near the pointing center (the star and inner disk). The peak intensity in the cleaned data is $1.74 \rm~ mJy~ beam^{-1}$, corresponding to a S/N of $\sim 25$. At $870 \mu m$ the thermal emission from the host star's photosphere contributes $<1\%$ to the peak flux per beam, assuming a blackbody with T$_{\rm{Eff}} = 10,500$ K, a radius of $1.7 ~ \rm R_{\odot}$, and a distance of $116$ pc.  The star's flux is thus negligible, as long as corona and chromospheric effects can be ignored.

The dust distribution is constrained using CASA's {\it uvmodelfit}, which fits single component models directly to the visibility data and selects the best fit through $\chi^{2}$ minimization. We run this task to fit a uniform disk model to the continuum data (the CO channels are {\it split} out) and list the best-fit model in Table \ref{UV_tab}. Disks with inclinations near $i\sim 55^\circ$ are favored with a major axis of about 0.''85, corresponding to $\sim 85$ au at a distance of $116~\rm pc$. The preferred model has a total continuum flux density of $3.78 \pm 0.23 ~\rm mJy$. This is within $15\%$ of the flux density found by summing the total flux from the cleaned image down to the $3\sigma$ contour.  The uncertainty in the flux is dominated by the uncertainty in the absolute flux scale, which is taken to be $10\%$.   This sets our flux estimate of the inner dust disk to be $3.8 \pm 0.4 ~\rm mJy$.

\begin{table}[H]
\caption{Summary of CASA's {\it uvmodelfit} results for the debris disk. The data were fit by comparing a simple, uniform disk model to the data visibilities. The fitting uncertainties for parameters other than flux are not included here, but are addressed for the gaseous disk in  section \ref{MCMC}. } 

\centering %
\begin{tabular}{c | c } 
\hline\hline 

    Parameter & Continuum [Debris]  \\
	\hline
    Flux Density & $3.78 \pm 0.23 ~\rm mJy$  \\
    X Offset & $-0."032$   \\
    Y Offset & -$0."023$  \\
    Major Axis & $0."85$  \\
    Axis Ratio (inclination) & $0.58$ [$55^{\circ}$]  \\
   Position Angle & $-8.8^{\circ}$

\label{UV_tab}
\end{tabular}
\end{table}

\subsection{Gas Disk}

\begin{figure}[H]
\centering
\includegraphics[width=.75\textwidth]{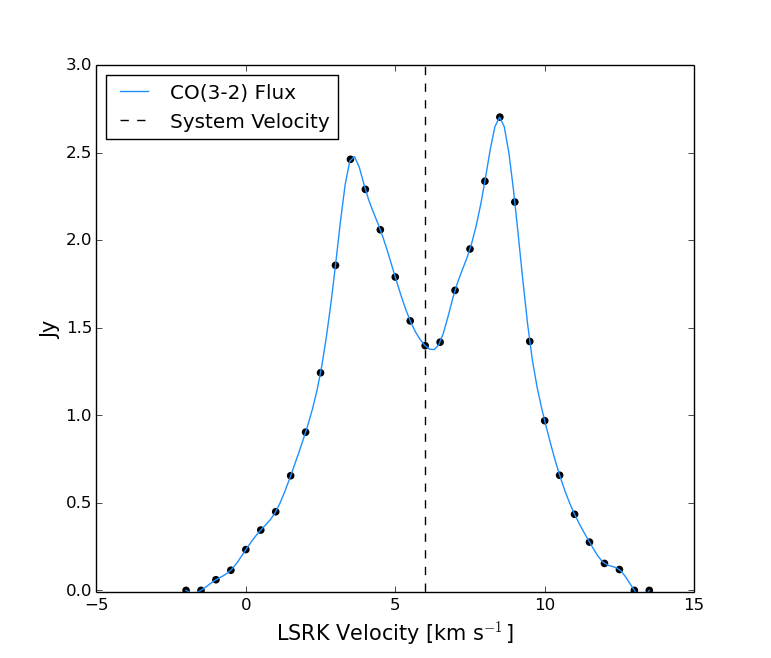}
\caption{Continuum subtracted CO(3-2) spectra as a function of LSRK velocity. The dashed line represents the system velocity of $6~\rm km~s^{-1}$. The $\sigma_{\rm RMS}$ of the individual channels is $\sim 6$ mJy meaning that the dominant source of uncertainty will come from the absolute flux calibration, which we take to be $\sim 10\%$. 
\label{CO}}
\end{figure}

In addition to the continuum, CO(3-2) emission is kinematically and spatially resolved using the FDM capabilities of the ALMA correlators, with a spectral resolution of $0.85~\rm km~s^{-1}$. The double-horned spectrum is shown as a function of LSRK velocity in Figure \ref{CO}. The previously constrained system velocity of $6~\rm km~s^{-1}$ is shown, as well as the asymmetric emission from the disk (Dent et al.~2005).  

\begin{figure}[H]
\centering
\includegraphics[width=1.1\textwidth]{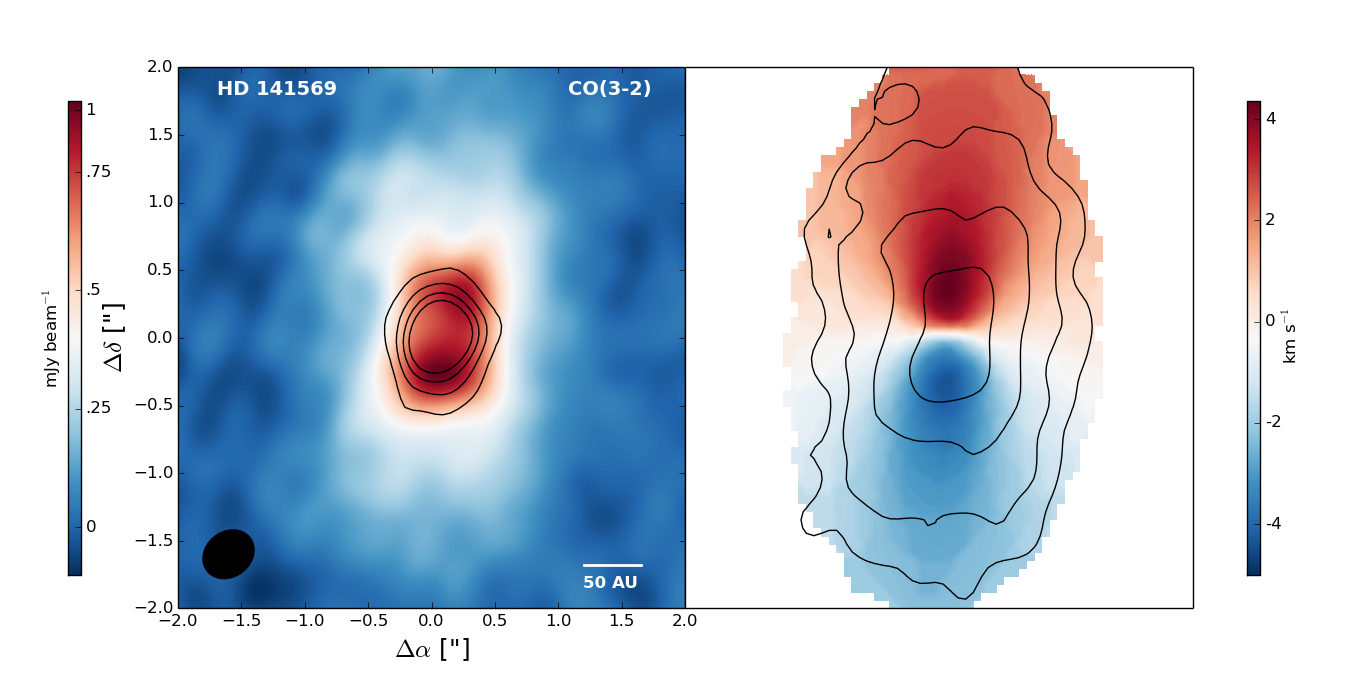}
\caption{\textbf{Left:} CO zeroth moment map. The contours represent $3, 6, 9 $ and $ 12 \times \sigma_{RMS}$ noise of the continuum ($\sigma_{RMS} = 0.070 ~\rm mJy~beam^{-1}$). The solid ellipse in the bottom left represents the beam size with properties as given in Table 2. A $50$ au scale (assuming a system distance of $116$ pc) is given in the bottom right. \textbf{Right:} CO first moment map (velocity field). The contours represent $3, 6, 12 $ and $ 24 \times \sigma_{RMS}$ noise ($\sigma_{RMS} = 0.028 ~\rm Jy~beam^{-1}$). Coordinates are given as offset from the phase center, as indicated on the left plot. North is up and East is to the left.
\label{moms}}
\end{figure}

The CO is continuum subtracted using the CASA task {\it uvcontsub}. Figure \ref{moms} (left panel) shows the brightness map for the CO line (zeroth moment), in which the $3\sigma$ CO contour extends out to $1.8"$ ($\sim 210$ au). This is compared directly with the continuum emission (contours), which is more centrally concentrated. The right panel shows the velocity map (first moment), with the CO brightness contours overlaid. There are two brightness peaks, each at about $\sim 0.9\rm~Jy~km~s^{-1}$. The peaks are separated by $\sim~0."5$ in a morphology that resembles ring ansae and suggestive of an inner gas cavity. There is only a tenuous CO detection within this $\sim~0."5$ ($\sim50\rm~au$) diameter cavity which is broadly consistent with previous shorter wavelength observations that find only tenuous CO between about 10 and 50 au in radius \citep{brittain02, goto}.

The velocity field map shows clear Keplerian rotation, with the gas south of the star approaching us. The brightness is skewed westward (right in the image), relative to the velocity map, which is discussed in more detail below below. 

\begin{figure}
\centering
\includegraphics[width=1.1\textwidth]{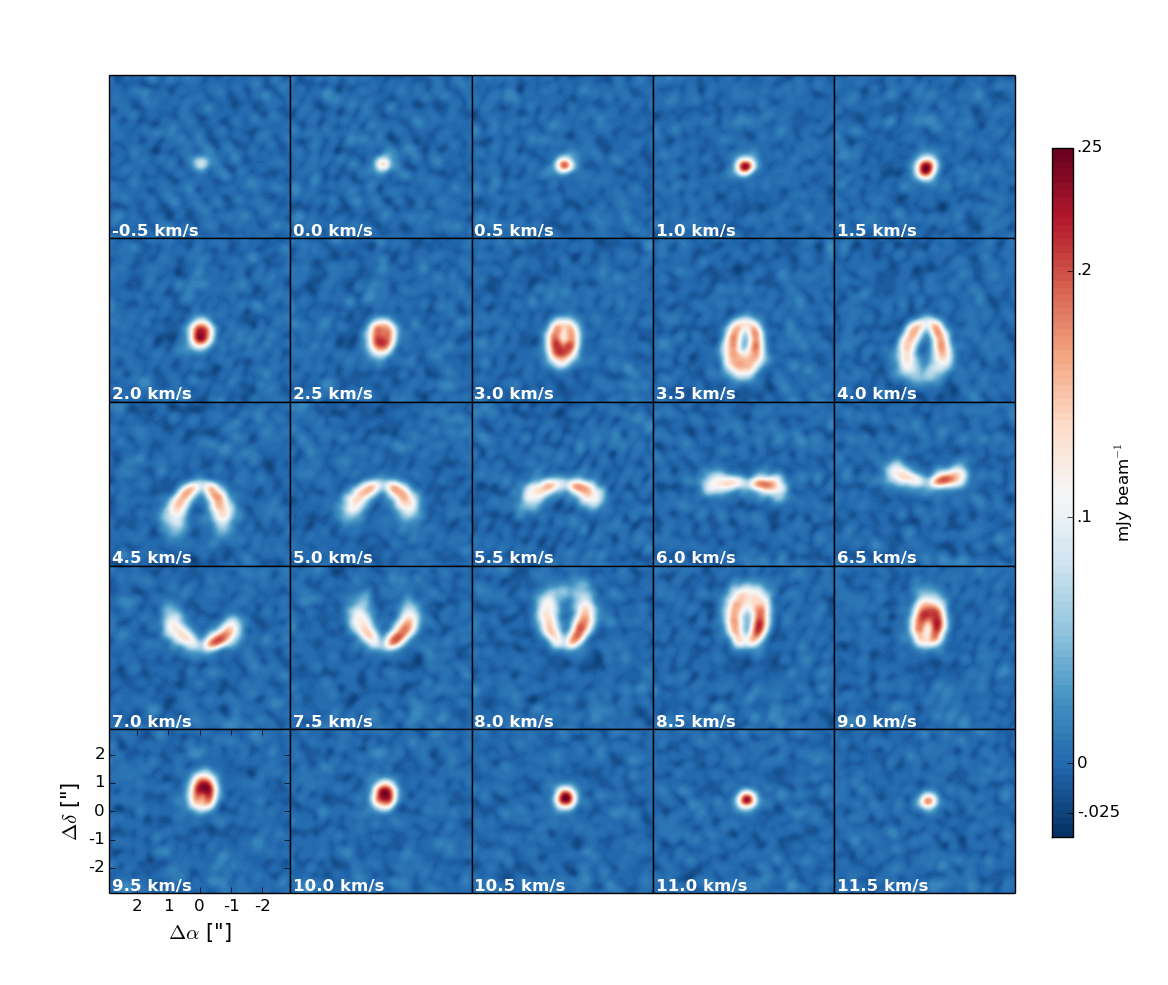}
\caption{ Channel map of the CO(3-2). The 25 subplots step forward in $0.5~\rm km~s^{-1}$ intervals from $-0.5$ to $11.5~\rm km~s^{-1}$ LSRK. The contours represent 3, 9, and 24 times the RMS noise of the intensity weighted map (as seen in Fig.\ref{moms}). Coordinates are given as offset from the phase center, as indicated on the bottom left plot. North is up and East is to the left. \label{chan}}
\end{figure}

Fig.\,\ref{chan} shows maps for 25 velocity channels between $-0.5\rm~km~s^{-1}$ and $11.5\rm~km~s^{-1}$. Contours represent 3, 6, 9, and 24 times the RMS noise of the zeroth moment. The spectral resolution of the velocity, $0.85~\rm km~s^{-1}$, is a factor of 2 larger than the channel width. For  Fig.\,\ref{chan}, velocity channel spacing is chosen to be $0.50~\rm km~s^{-1}$ to include a slight oversampling. The total flux density of the CO given in Table \ref{data} is determined by summing the flux in the zeroth moment map down to $3 \times \sigma_{RMS}$ and multiplying by the number of beams. This value is consistent with integrating over all channels of the CO map to within $10\%$. Note again that there is a clear asymmetry in the emission west of the star.

The peak flux in the northwestern limb is significantly brighter than its counterpart in the northeastern and southwestern limbs.

\subsection{MCMC Modeling}

As shown in Figure \ref{moms}, the high spatial and velocity resolution capabilities of ALMA yield a well-constrained velocity field. These data can thus be compared with a Keplerian disk model to infer system properties.  Trial models are generated by first assuming a uniform Keplerian disk.  For simplicity, the inner cavity, temperature profile, and line broadening of CO (which is expected to be small) are not factored in to the model. Each model is projected to the disk geometry and the LSRK velocity is subtracted. The model is then convolved with a 2D Gaussian beam as given in Table 2. Using Markov Chain Monte Carlo (MCMC) techniques (specifically, Metropolis-Hastings with Gibbs sampling), the posterior distributions are calculated for the disk's inclination, position angle, LSRK system velocity, dynamical center, and mass. We assume flat prior distributions over the ranges given in Table \ref{mcmc_pri}. Model comparison is conducted in the image domain due to the high velocity resolution and signal-to-noise.

\begin{table}[H]
\caption{Ranges for the flat prior distributions of each parameter.  The Gaussian widths are also given for the proposal distributions. The prior is based on the UV model fitting results given in Table \ref{UV_tab}. }
\centering 
\begin{tabular}{c | c | c } 
\hline\hline

    Parameter & Prior Range & $\sigma$ \\  
	\hline
	Mass [M$_{\odot}$]     & $[1.0, 4.0]$   & 0.02  \\
	Position Angle [$^{\circ}$]   &  $[-15.0, 5.0]$ & 0.1 \\
	Inclination [$^{\circ}$]      &  $[45.0, 65.0]$  & 0.2\\
	System Velocity [km s$^{-1}$] & $[5.0, 7.0]$   & 0.01 \\
	X Offset ["]  &  $[-0.2, 0.2]$ & 0.06 \\
    Y Offset ["]  & $[-0.2, 0.2]$ & 0.06  \

\label{mcmc_pri}
\end{tabular}
\end{table}

Parameter space is explored through a random walk directed by Metropolis-Hastings MCMC  \citep[e.g.,][]{ford}. For each new trial, two model parameters are randomly chosen and then updated by drawing a Gaussian random parameter centered on the current model (state $i$). The acceptance probability for the new trial model (state $i+1$) is given by 
\begin{equation}
\alpha =\rm{min}(e^{\frac{1}{2}\left(\chi^{2}_{i}-\chi^2_{i+1}\right)},1),
\end{equation}
where we take
\begin{equation}
\chi^{2}_{i} = \sum{ \frac{(D - M_{i})^{2}}{\sigma^{2}}}.
\end{equation}
Here, D are the data from the CO first moment map (see Fig.\,\ref{CO}), M$_{i}$ is the current model, and $\sigma = 0.5 ~\rm km~s^{-1}$ is the velocity channel width.  The summation is over all points on the moment map. If $\alpha$ is greater than a random number drawn from a uniform [0,1] distribution, then the new model is accepted and recorded in the Markov chain. If the model is rejected, then the previous model is used again and re-recorded.

\begin{figure}[H]
\centering
\includegraphics[width=\textwidth]{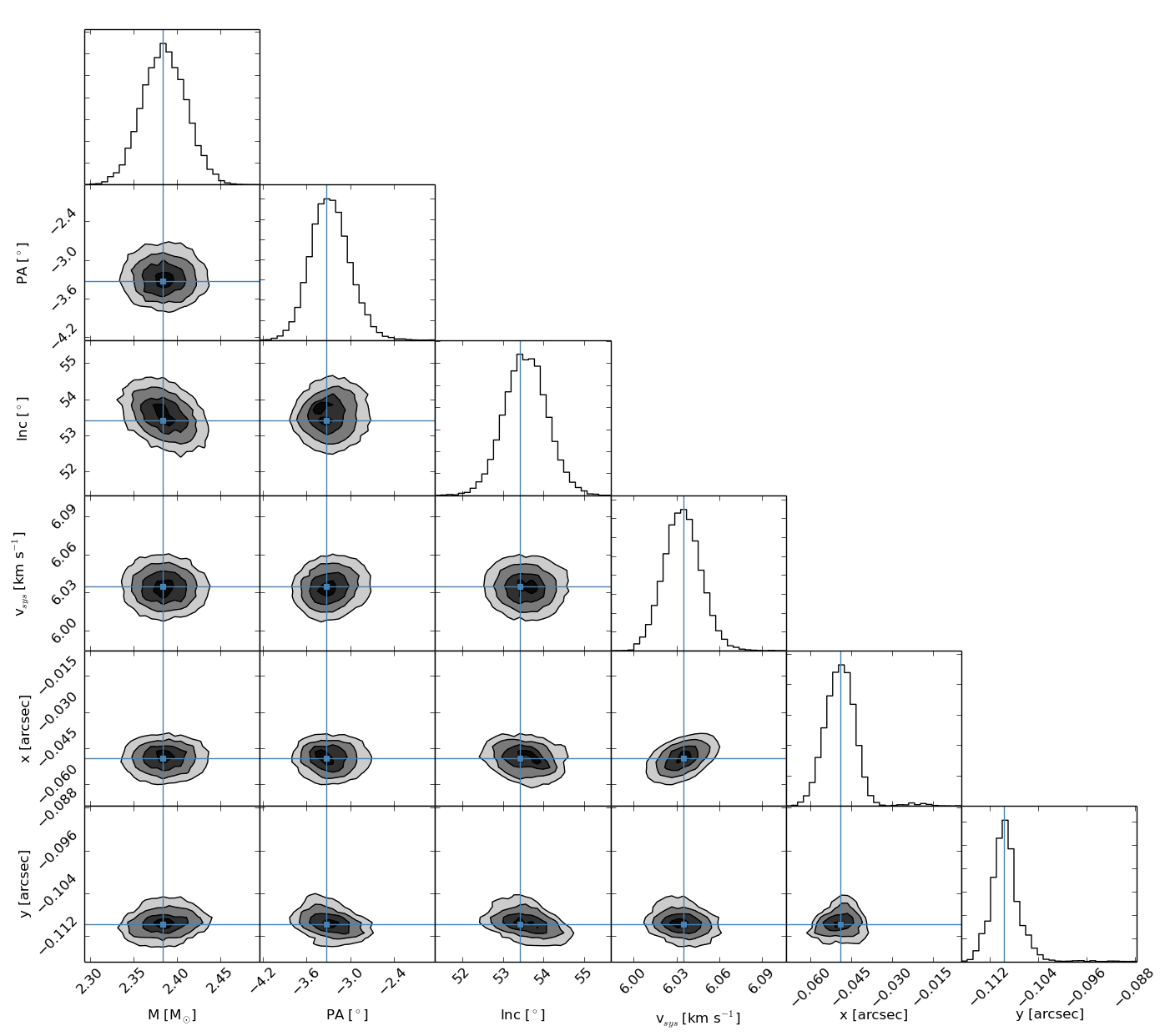}
\caption{Posterior probability distribution from MCMC modeling of the CO velocity field for 300 thousand links minus the burn-in. The blue points represent the most probable model parameter. The contours show 0.5, 1, 1.5, and 2$\sigma$.
\label{MCMC}}
\end{figure}

The MCMC routine is run using 3 chains, each with randomly chosen starting points in the flat prior parameter space.  Each chain contains 100 thousand links of which about 1000 are needed for burn-in. The 3 chains converge on similar parameters, and the distributions are combined to give the resulting posterior distributions in Fig.~\ref{MCMC}. The blue points correspond to the values of highest probability. The most probable parameters (i.e., the mode of the distributions) are given in Table \ref{mcmc_par}. Uncertainties are given by a $95\%$ credible interval unless otherwise stated. The most probable mass is $2.39~\rm M_{\odot}$, for a distance\footnote{The most probable mass scales directly with the assumed distance.} of 116 pc. Since there is a degeneracy in inclination and mass, we give ${\rm M} \sin(i)$ and ${\rm M}$. Previously constrained stellar mass estimates are between 2.0 and 3.1 $\rm M_{\odot}$ \citep[e.g.,][]{merin,wyatt07}. The posterior distributions for both quantities are sampled independently by the MCMC. Ultimately, the uncertainty in the derived mass is dominated by the distance uncertainty. The re-analyzed {\it Hipparcos} Catalog distance with 1-$\sigma$ uncertainty is $116\pm8$ pc \citep{van}.  Considering only this 1-$\sigma$ distance uncertainty with our most probable mass yields $2.39^{+.16}_{-.16}\rm~M_{\odot}$.

\begin{figure}[H]
\centering
\includegraphics[width=\textwidth]{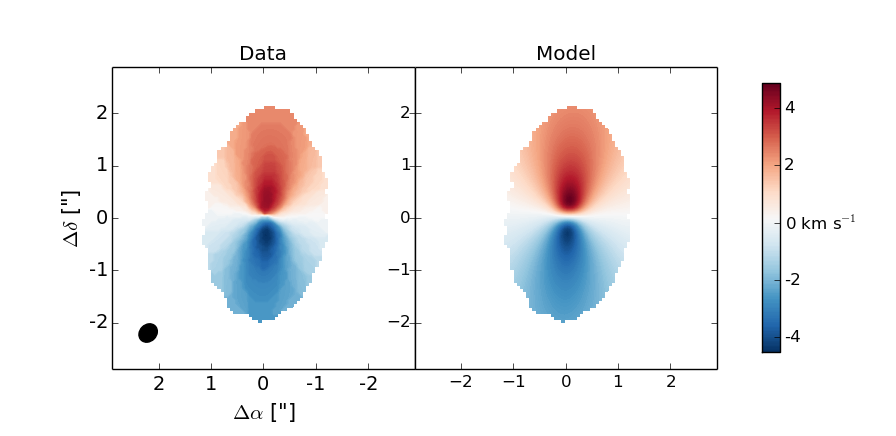}
\includegraphics[width=.8\textwidth]{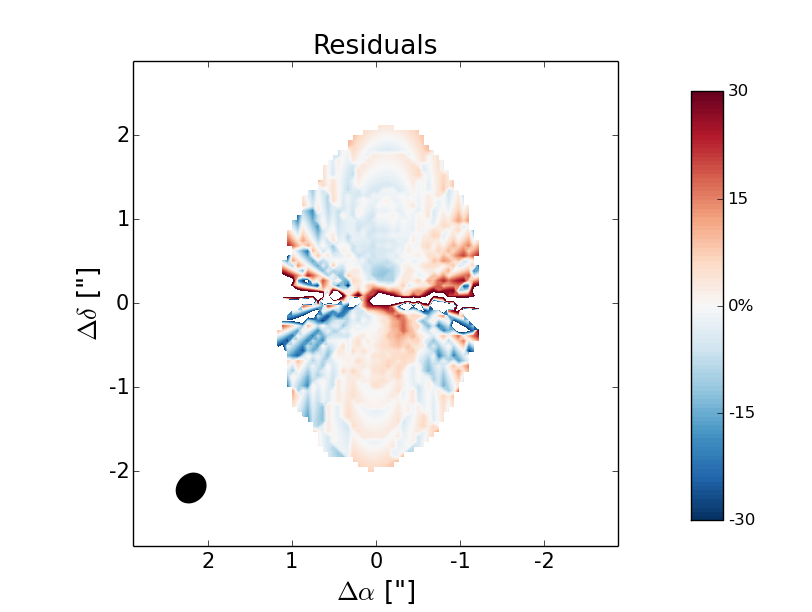}
\caption{\textbf{Top:} The left panel shows the first moment map of the data (same as RHS in Fig.\,\ref{moms}), while the right shows the velocity field of the model. \textbf{Bottom:} The panel shows the residuals presented as a percent difference in the model from the data. All images are shifted to the system centered velocity of $6.04~\rm km~s^{-1}$. The model is consistent with the data to about 10\% or better throughout most of the disk.  The largest deviations occur along the minor axis. The black ellipse in the bottom corresponds to the beam with properties given in Table 2.
\label{resid}}
\end{figure}

The most probable parameters are used to construct a final disk model, which is shown in Figure~\ref{resid}. The residuals of the model are also shown as percent deviation from the data. The most probable model typically shows agreement with the data to about 10\%, but has larger deviations along the minor axis of the data/model.

\begin{table}[H]
\caption{Summary of MCMC Results with $95\%$ Credible Range.} 
\centering 
\begin{tabular}{c | c | c} 
\hline\hline

    Parameter & Most Probable & $95\%$ Credible Range \\
	\hline
	Mass [M$_{\odot}$]     & $2.39$  & $[2.34,2.43]$ \\
	Mass [M$_{\odot}$sin(i)]     & $1.92$  & $[1.89,1.95]$ \\
	Position Angle [$^{\circ}$]   & -$3.36$ & $[-3.78,-2.71]$ \\
	Inclination [$^{\circ}$]      & $53.4$ & $[52.5, 54.6]$   \\
	System Velocity [km s$^{-1}$]& $6.04$  & $[6.01, 6.06]$     \\
	X Offset ["] & $-0.049$ &  $[-0.060, -0.038]$ \\
    Y Offset ["] & $-0.11$ & $[-0.12, -0.10]$ \

\label{mcmc_par}
\end{tabular}
\end{table}

\section{Discussion}

\subsection{Disk Asymmetry}

An interesting asymmetry is observed in the CO channel map. Looking at the ``butterfly'' features in the $4-8 \,{\rm km~s^{-1}}$ channels, there is a localized flux enhancement on the northwestern (top right) component of the gas. The east wing of the butterfly has a fairly symmetrical intensity about the system velocity of $6 \,{\rm km~s^{-1}}$, while the west wing is asymmetrical.

\begin{figure}[H]
\centering
\includegraphics[width=\textwidth]{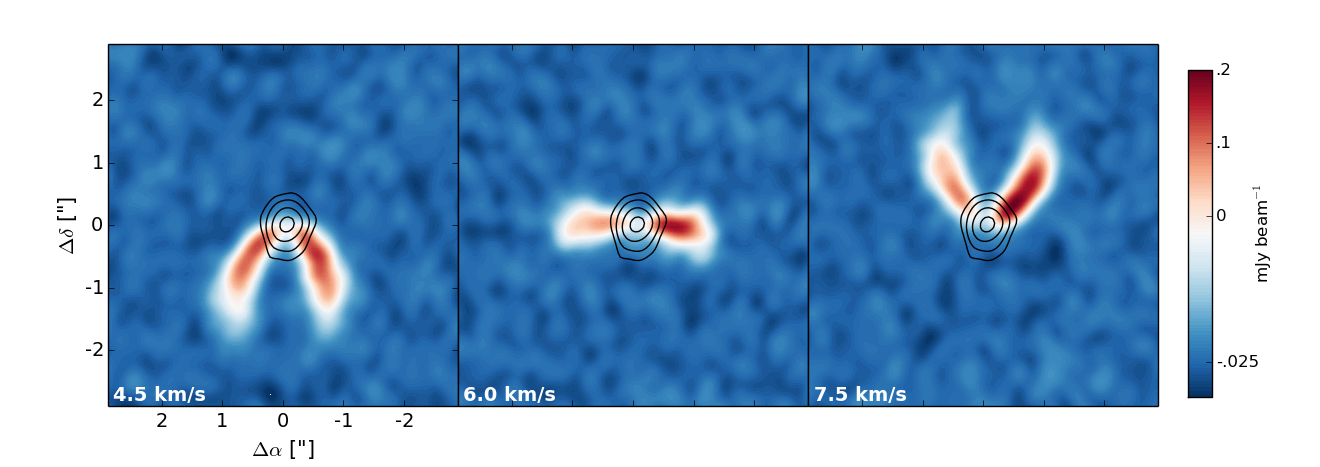}
\caption{The $4.5, 6.0,$ and $7.5  ~\rm km~s^{-1}$ velocity channels of CO. The localized flux enhancement can be seen on the west and northwest components of the gas disk. The velocities given are LSRK and are centered around a system velocity of $6  ~\rm km~s^{-1}$. The contours represent $3, 6, 12 $ and $ 21 \times \sigma_{RMS}$ noise ($\sigma_{RMS} = 0.070~\rm mJy~beam^{-1}$) fo the continuum. North is up and East is to the left.
\label{zoom}}
\end{figure}

To explore this feature further, Figure \ref{zoom} shows three of the channel maps (4.5, 6, and 7.5 ${\rm km~s^{-1}}$), along with the continuum using $3, 6, 9 $ and $ 12 \times \sigma_{rms}$ contours. The southern components of both sides appear to be approximately symmetric, but a strong asymmetry becomes obvious for the 6 and 7.5 ${\rm km~s^{-1}}$ maps, in which the western wing is brighter than the eastern wing by $\sim 40\%$ in each channel. These channel maps also suggest that there is indeed an inner cavity to the CO disk, as noted in other studies \citep{goto, flaherty}.

Since the asymmetry is present throughout multiple channels (see Fig.\,\ref{chan}), the feature appears to be real in the data. While the exact source of the flux enhancement is unknown, it may be caused by asymmetries in the inner disk edge, such as vortex formation \citep[e.g.,][]{lyraa,lyrab} or by perturbations from an unseen companion. \citet{dent} observe a large asymmetry in $\beta$-Pic that is attributed to localized collisions of gas-rich comets.  The asymmetry is also in the general direction of the two distant red dwarf companions that orbit at $\sim 1000 ~ \rm au$. Follow-up observations and detailed simulations are required to determine the cause of the CO disk morphology.

\subsection{Debris/Dust Mass \label{sec:debris}}

An initial estimate for the dust mass is made by assuming that the emission is optically thin, dominated by mm grains, and spatially concentrated in a thin ring. In this case,
\begin{equation}
M = \frac{4}{3} \rho_i \pi s^{3} \frac{F_\nu(\text{Obs})}{ B_\nu(R)\Omega_{\text{s}}},
\end{equation}
where $F_\nu \text(Obs)$ is observed  flux density of the continuum, $B_\nu {(R)}$ is the black body intensity for a single grain placed at a distance $R$ from the star, and $\Omega_s$ is the solid angle of a single grain.  The grains are further assumed to be in thermal equilibrium with the host star, to have an internal density $\rho_i = 2.5~\rm g~cm^{-3}$ and size $s=1~\rm mm$, and to be perfect absorbers and radiators (albedo of 0, emissivity of 1). We note that this mass estimate  is equivalent  to  $M  = \frac{d^2 F_\nu(\text Obs)}{\kappa_\nu B_\nu {(R)}}$ with $\kappa_\nu =3~\rm cm^2~g^{-1}$, for our assumptions.  This opacity is within a factor of two of the mm opacity used by \cite{flaherty}. 

For an approximate lower limit, the ring can be envisaged to be at $R=10$ au, which represents the innermost location for large grains based on SED modeling \citep{malfait}.
At this distance and for the noted assumptions, the grains would be  $T\sim200~K$, which yields a mm grain mass\footnote{Adopting a different distance will scale the mass by $(\frac{d}{116~\rm pc})^{2}$} of $0.04~\rm M_{\oplus}$.   Placing grains at larger stellar separations would require additional mass to explain the emission. For example, if all the grains were placed at $R=50$ au ($T\sim90~K$), the mm grain mass would be $\sim0.09~\rm M_{\oplus}$. 

This simple estimate may only correspond to the actual dust mass if the observed mm grains are leftovers that were never incorporated into planets.  Instead, if the grains are produced by the evolution of a nascent debris disk, the total mass can be significantly different. We explore this possibility next using a size distribution of grains spread throughout a disk.

For simplicity, we assume that the surface density of material decreases as $r^{-1}$ over disk radii 10-50 au. The dust is assumed to be absent outside of these boundaries. We further take the grains to radiate efficiently as long as the their diameter $(2s)$ is equal to or larger than the absorbing/emitted photons \citep[e.g.,][]{wyatt}. For wavelengths larger than the grain's diameter, the emission and absorption coefficients ($\rm Q_\nu({\rm em}) = Q_\nu ({\rm abs}) = Q_\nu$) are inversely proportional to the photon wavelength.  Specifically,
\begin{equation}
 {\rm Q}_\nu =  \left\{
  \begin{array}{lr}
    1 &  ~ 2s > \lambda\\
    \frac{2s}{\lambda} &  ~\rm otherwise.
  \end{array}
\right.
\end{equation}

We only consider a ``total'' debris mass up to some maximum parent body size, which is taken to be $s_{\rm max}=50\rm~km$. This does not mean that $50~\rm km$ is envisaged to be the largest solids in the debris disk; it is only the maximum size we consider in a given size distribution. To get a total debris mass for solids $s<50$ km, a particle size distribution must be assumed. Lacking further constraints, we use a collisional cascade such that the mass per size increment $M_s\propto s^{-0.5}$ \citep[e.g.,][]{dohnayi}. The total mass is then determined by requiring the model continuum flux density to match our observations. In practice, the debris disk is divided into a series of rings (here 100), placed evenly between 10 and 50 au.  If each ring has the same mass, then the surface density profile follows $r^{-1}$. A flux density for each ring is then calculated by first deriving a grain temperature, assuming that the grains are dark (albedo$\sim0$) and balancing the received and emitted powers using a black body model with the effects of $Q_\nu$. The grain temperature \citep[e.g.,][]{wyatt} is

 \begin{equation}
 T_g = T_{g,BB} \left(\frac{Q_{\rm abs}(T_\text{star})}{Q_{\rm abs}(T_g)}\right)^{1/4}, 
 \end{equation}
 
 where $T_\text{star}$ is the host star's surface temperature (assuming it is a black body), $T_{g,BB}$ is the equilibrium grain temperature if the grain were also a perfect black body, and $T_g$ is the actual grain temperature. The equation must be solved iteratively, but converges quickly.  For HD141569, $T_g\approx T_{g,BB}$ except at grains less than 10s of microns. To calculate $Q_{\rm abs}(T)$,  $Q_\nu$ is integrated over all frequencies and weighted by a black body of the given temperature, i.e., 
 \begin{equation}
 Q_{abs}(T) = \frac{ \int B_{\nu}(T,\nu) Q_\nu  d\nu}{ \int B_{\nu}(T,\nu) d\nu}.
\end{equation}
Taking $T_\text{star} = 10500$ K and the above grain size and spatial distribution, we find that $M(s<{\rm 50~km})\sim 160~M_{\oplus}\frac{\rho_i}{\rm 2.5~g~cm^{-3}}$. 

This result should be interpreted with caution.  A steeper (shallower) solid size distribution can lead to significantly larger (smaller) masses. The result is also dependent on the internal density of the grains, as well as their effective albedo and emissivity. Nonetheless, the results are illustrative that significant debris may be distributed between 10 and 50 au. The total mass of solids would be much larger should debris (at a lower surface brightness) be present at disk radii $r>50$ au, which would be consistent with single dish measurements (see Table 1).

\subsection{Gas Mass}

The mass of an optically thin gas disk near LTE can be calculated from the integrated line intensity \citep[e.g.,][]{perez}. Given a line flux of F$_{\rm{OBS}} = 15.7 \rm~Jy~km~s^{-1}$, the average line intensity over the source's solid angle $\Omega$ is
\begin{equation}
\hat{I} = \frac{F_{\rm OBS}}{\lambda \Omega} ,
\end{equation}
where $\lambda = 867 \mu$m is the average wavelength of the observations. The upper transition level column density of CO is given by
\begin{equation}
N_{3} = \frac{4 \pi \hat{I}}{h \nu A_{32}} ,
\end{equation}
where $\nu = 345.79$ GHz is the frequency of the molecular feature, and $A_{32} = 2.497 \times 10^{-6}$ Hz is the Einstein absorption coefficient\footnote{The spectral information for the CO molecule was obtained from the Splatalogue database http://www.splatalogue.net, \citet{splat}.} for the transition. 

In the following, $J=3$ (the upper transition level) unless otherwise noted (such as in the summation). Under the assumption that all $J$ energy levels are populated in LTE, the total column density is given by
\begin{equation}
N_{\rm{Total}} = N_{J} \frac{Z}{2J+1} e ^{\frac{h B_{e} J(J+1)}{kT}},
\end{equation}
and Z is
\begin{equation}
Z = \sum_{j=0}^{\infty} (2j+1) e ^{-\frac{h B_{e} j(j+1)}{kT}}.
\end{equation}
Here, $B_{e} = 57.635$ s$^{-1}$ is the rotational constant$^{4}$, T is the gas temperature, Z is the canonical partition function. The gas mass is then given by
\begin{eqnarray}
\rm{M}_{\rm{CO}} &=&  m_{\rm{CO}}  N_{\rm{Total}} \Omega d^{2},\\\nonumber
 & = &  \frac{4 \pi m_{\rm{CO}} \, d^{2} \,F_{\rm OBS} \, \rm{Z}}{h \nu \lambda  A_{32} (2J+1)} ~ e ^{\frac{h B_{e} J(J+1)}{kT}}
\end{eqnarray}
for a solid angle $\Omega$ and distance to the object $d$.
Taking a gas temperature of T $ = 33$K, the minimum excitation temperature of the  $J= 3$-2 line, gives M$_{\rm{CO}} =1.9 \pm 0.2 \times 10^{-3}~\rm M_{\bigoplus}$, with the uncertainty propagated from the CO flux density  uncertainty in Table 2. The corresponding spatially averaged column density, $N_{\rm{Total}}$, is $1.2 \pm 0.1 \times 10^{16} ~\rm cm^{-2}$. This is within the optically thin limit \citep{wyatt_stages}, but should not be taken as independent confirmation, as we assumed the gas to be thin for the mass calculation. If the gas is partly optically thick, then the actual CO gas mass could be larger by a factor of a few \citep{matra}. As such, the CO mass here could be interpreted as a lower limit. Due to the uncertainty in the appropriate amount of the gas, we will only report the CO mass as M$_{\rm(CO)}\sim2\times10^{-3}~\rm M_{\bigoplus}$ to emphasize that the calculation has important unknowns. 

\citet{flaherty} find a gas model with total mass of $13^{+50}_{-9}~\rm M_{\bigoplus}$ as constrained by LTE models of gas temperature and density of CO(1-0) and CO(3-2) with CARMA and SMA, respectively. If we assume the $10^4$ ISM number density abundance ratio for H$_{2}$ to CO  (as in Flaherty et al.), the inferred H$_2$ gas mass from the ALMA observations is  M$_{\rm{H_2}} \sim 1.4 ~\rm M_{\bigoplus}$. Including additional metals would increase the total inferred gas mass to be slightly above $\sim 1.5 ~\rm M_{\bigoplus}$, which is a factor of a few below the lower bound of the SMA and CARMA based model. The observations and models altogether thus suggest that there is 1 to a few tens M$_\oplus$ of gas mass, assuming the ISM scaling can be used, which is not obviously the case. Additional caveats for these gas-mass estimates are discussed below.

\subsection{What can HD141569 tell us about grain growth, planet formation, and disk evolution?}

The morphology of HD141569 shows a dust disk extending out to about 56 au and an extended CO gas component between about 30 and 210 au.  This structure alone suggests that the system is an evolved transition disk. However, as discussed below, HD141569 may be better interpreted as a nascent debris system. The distinction is that the dust would be second generation, and any associated size distribution would reflect the clearing stages of planet formation rather than grain growth outcomes.

\subsubsection{Primordial v.~Second Generation}
 
While most debris disks are expected to be extremely gas-poor, several younger debris systems (e.g., $\beta$ Pic as discussed in the Section 1) have been observed with CO masses $ \rm M_{\rm CO} = 10^{-5} - 10^{-2} ~\rm M_{\bigoplus}$ \citep{pascucci, hughesa, dent}. HD141569 has a CO gas mass $\sim 2\times10^{-3} \rm M_\oplus$, which, while younger, is comparable to these more evolved systems. The total gas mass of  $\sim1.5~\rm M_{\bigoplus}$ ($\sim 5 \times 10^{-3}\rm M_J$) assumes an ISM H$_2$ to CO abundance ratio.  There is ultimately no reason to suspect that this conversion is applicable to HD141569 after 5 Myr of evolution. If the gas disk is not optically thin, as assumed in the calculation above, then using CO as a tracer of total gas could underestimate the actual gas mass \citep{bergin}.

The current CO disk should be expected to be depleted by photodissociation on timescales of $\sim120$ yr \citep{visser}, unless significant self-shielding is present. While the derived column density of CO  ($\sim 10^{16}\rm~cm^2$) would contribute to some shielding, it is not obviously sufficient to prevent rapid dissociation. Unless the gaseous disk is massive enough to prompt CO formation in rough balance with photodissociation, the low inferred CO mass creates a potentially serious timing problem for a primordial gas interpretation. 
Instead, if the gas is second-generation as produced by a planetesimal population  \citep[e.g.,][]{moor11}, then the short dissociation timescale may not be problematic.  Rather, the problem now becomes whether sufficient mass is available to produce a low-mass gaseous disk, and if so, whether the planetesimal destruction rates would be consistent with  the dynamics and the radiation field of the system.  

First, we note that the debris interpretation is corroborated by recent scattered light imaging \citep{konishi}. The images reveal very small grains present around 50 au, a region co-located with the mm grains observed here.  Such small grains should be removed by the system quickly by radiation pressure.  The presence of the small dust grains in this region of the disk suggests that significant collisional evolution is indeed taking place. This, by itself, does not suggest that the gaseous disk is best described by a debris disk, but it motivates its consideration.

If the CO gas is depleted quickly through photodissociation on $\sim120$ yr timescales, then for our estimate of the CO mass, the CO production rate must be $\rm \dot{M}_{\rm CO}\approx 17 ~  M_\oplus ~ Myr^{-1}$.
If a typical comet's mass is 10\% CO ice \citep{mumma}, then about 170 $\rm M_\oplus$ of cometary material must be destroyed per Myr to balance photodissociation.  This also implies that the total gas mass is within an order of magnitude of the CO gas.  Based on cometary compositions, CO can be  accompanied by approximately similar abundances of H$_{2}$O and CO$_{2}$ \citep{mumma}. Ultimately, spectroscopic followup must be used to determine the gas composition and compare that with cometary abundances to further constrain this scenario observationally.  

Is the required comet destruction rate plausible?  As discussed in Section \ref{sec:debris}, the ALMA continuum emission of 3.8 mJy with a collisional cascade model implies a total solid mass $M\sim160~\rm M_\oplus$ for $s<50~\rm km$ in the inner disk.  While the ALMA CO observations are consistent with single-dish observations, the continuum flux measured here is lower than that found in previous studies.  For example, single dish observations by \cite{nilsson} find a continuum flux density of  $12.6\pm 4.6~ \rm mJy$  at 870 $\mu m$, and the SMA observations measure $8.2\pm 2.4~ \rm mJy$ \citep{flaherty}. The much larger beam in these observations could be biasing the detected flux through contamination, but at face value, this suggests that there may still be considerable dust mass at larger radii whose emission is resolved out by the interferometer or is too low surface brightness to be detected at the sensitivity of these observations. As such, the true mass in solids may be larger than estimated here. For example, if we extend the collisional cascade model out to 210 au (the extent of the CO disk) and normalize the mass to 12.6 mJy, the total solid mass is over 360 $\frac{\rho_i}{\rm 1~g~cm^{-3}}\rm M_\oplus$ (for $s<50$ km), where we have used $\rho_i= 1\rm~g~cm^{-3}$ to represent icy bodies. We stress that this estimate is very uncertain, as it depends on the assumed size distribution, planetesimal densities, grain albedos and emissivities, and distance to HD141569\footnote{This estimated debris mass for an extended disk scales as roughly $(\frac{d}{\rm 116~pc})^2$.}.  Provided that the estimated mass reservoir is dynamically accessible (which is not explored here or obviously met), there is potentially sufficient cometary material to produce the current CO gas, although the system would not maintain this gas abundance for a protracted time without shielding.

Why should significant CO gas only appear outside a radius of about 30 au? As noted in the introduction, tenuous, warm CO has been detected interior to the 50 au diameter cavity, but there is a large change in CO abundance exterior to this distance, as revealed here. If the gas is indeed second generation, then the change in CO abundance may reflect where significant CO was incorporated into planetesimals at the time of their formation.  In this paradigm, the entire disk is collisionally evolving, but significant CO gas is only released in planetesimals that harbor a large fraction of CO ice.  Alternatively, the reduced abundance of CO interior to about 30 au may simply reflect the CO photodissociation environment closer to the star and/or changes in self-shielding. The inner edge of the CO could also be set by a region with a higher rate of stirring by planets and embryos \citep{lissauer}.

There is a potential contradiction, however, with this approach.  The CO mass was derived assuming that it is optically thin and in LTE.  If the gas is indeed second-generation, then it is not obvious whether there will be sufficient collisional partners to populate the rotational levels thermally.  In this case, the true CO mass could be significantly different from our estimates, and potentially even orders of magnitude more massive if non-LTE effects do dominate \citep{matra}.  To check the degree to which the LTE assumption may be valid, we use the ALMA measured CO(3-2) integrated line flux to estimate the CO(1-0) integrated line flux under LTE conditions, which is approximately $\sim 0.8\rm~Jy~km~s^{-1}$.  The \cite{flaherty} CARMA observations found an integrated line flux for CO(1-0) of $1.6\pm 0.2\rm~Jy~km~s^{-1}$, making the estimate good to about a factor of two.  
Ultimately, observations of disk chemistry are needed to understand the gas's origin.

\section{Summary}

We have presented ALMA continuum ($870 \mu m$) and CO(3-2) observations of HD141569. The continuum observations show a  dust disk that extends out to  $0."49$ with a total continuum flux density of $3.8 \pm 0.4 ~\rm mJy$ (peak flux of $1.74 \pm 0.24 ~\rm mJy~beam^{-1}$).  A rough lower limit to the amount of dust mass needed to explain the emission is $0.04~\rm M_{\bigoplus}$.  If the dust is due to the collisional evolution of debris (rather than leftover millimeter grains from planet-building), then the millimeter flux reflects a comet and asteroid reservoir of  $\sim 160~\rm M_{\bigoplus}$ for sizes $s<\rm 50~ km$ (assuming a collisional cascade).  The continuum flux density found here is about a factor of three lower than that derived by single dish observations, suggesting that there is additional dust on larger spatial scales or at a lower surface brightness.  

The CO disk observations reveal CO extending from roughly the outer edge of the inner dust disk to about $1."8$.  The CO(3-2) integrated flux density is $15.7 \pm 1.6~\rm Jy~km~s^{-1}$ (peak flux of $0.90 \pm 0.16 ~\rm Jy~km~s^{-1}~beam^{-1}$), which is consistent with single dish measurements.  Assuming that the gas is in LTE and optically thin, the corresponding CO mass is $\sim2 \times 10^{-3}~\rm M_{\bigoplus}$ for a distance of 116 pc.  

Based on modeling the velocity field, the disk is constrained to have a Position Angle $=-3.36^{\circ +.65}_{-.42}$, an inclination $=53.4^{\circ +1.2}_{-.9}$, and a system velocity $v_{sys} = 6.04^{+.02}_{-.03}$ km s$^{-1}$.  The gas velocities are consistent with orbiting a star of $2.39^{+.04}_{-.05}\rm~M_{\odot}$ for the most probably inclination and a distance of 116 pc. The uncertainties represent the 95\% confidence region computed from MCMC samples. Instead, considering only the 1-$\sigma$ distance uncertainty with our most probable mass yields $2.39^{+.16}_{-.16}\rm~M_{\odot}$.

The channel maps show a localized flux enhancement of the disk to the western section of the disk. Further detailed modeling of the system and higher resolution imaging are needed to properly constrain the full morphology.  Because CO should photodissociate rapidly, the gas may require, in part, replenishment through collisions of comets, making the disk a debris system.  While the required mass to do this may be high, it is potentially within plausible limits of the inferred debris field.  Observations probing the gas composition can be used to further constrain the origin of the gas, particularly as LTE assumptions may not apply.

We thank the anonymous referee for the helpful comments during the review process. J.A.W. and A.C.B acknowledge support from an NSERC Discovery Grant, the Canadian Foundation for Innovation, The University of British Columbia, and the European Research Council (agreement number 320620). A.M.H. and K.M.F. are supported by NSF grant AST-1412647. E.B.F.'s contribution was supported in part by funding from the Center for Exoplanets and Habitable Worlds.  The Center for Exoplanets and Habitable Worlds is supported by the Pennsylvania State University, the Eberly College of Science, and the Pennsylvania Space Grant Consortium. A.C.B. and E.B.F. also acknowledge The University of Florida and the NASA Sagan Fellowship program. M.J.P. also acknowledges NASA Origins of Solar Systems Program grant NNX13A124G, NASA Origins of Solar System Program grant NNX10AH40G via award agreement 1312645088477, NASA Solar System Observations grant NNX16AD69G, BSF Grant Number 2012384, as well as support from the Smithsonian 2015 CGPS/Pell Grant Program.

This paper makes use of the following ALMA data: ADS/JAO.ALMA[2012.1.00698.S] . ALMA is a partnership of ESO (representing its member states), NSF (USA) and NINS (Japan), together with NRC (Canada), NSC and ASIAA (Taiwan), and KASI (Republic of Korea), in cooperation with the Republic of Chile. The Joint ALMA Observatory is operated by ESO, AUI/NRAO and NAOJ.

\end{document}